\documentclass[conference]{IEEEtran}
\IEEEoverridecommandlockouts
\usepackage[table]{xcolor}

\usepackage{cite}
\usepackage{amsmath,amssymb,amsfonts}
\usepackage{algorithmic}
\usepackage{graphicx}
\usepackage{textcomp}
\usepackage{xcolor}

\usepackage{multirow,booktabs}
\usepackage{subcaption}

\makeatletter
\newcommand{\linebreakand}{%
  \end{@IEEEauthorhalign}
  \hfill\mbox{}\par
  \mbox{}\hfill\begin{@IEEEauthorhalign}
}
\makeatother

\newcommand{\tabincell}[2]{\begin{tabular}{@{}#1@{}}#2\end{tabular}}
\def\BibTeX{{\rm B\kern-.05em{\sc i\kern-.025em b}\kern-.08em
    T\kern-.1667em\lower.7ex\hbox{E}\kern-.125emX}}
\begin{document}

\title{LlamaPartialSpoof: An LLM-Driven Fake Speech Dataset Simulating Disinformation Generation

\thanks{
\textsuperscript{\dag}Dr. Hieu-Thi Luong is funded by RIE2025 NRF International Partnership Funding Initiative. This research is supported by the National Research Foundation, Singapore, under the AI Singapore Programme (AISG Award No.: AISG2-TC-2023-011-SGIL). Any opinions, findings and conclusions or recommendations expressed in this material are those of the author(s) and do not reflect the views of National Research Foundation, Singapore.\\ 

\textsuperscript{*}equal contribution}
}

\author{\IEEEauthorblockN{Hieu-Thi Luong\textnormal{*}}
\IEEEauthorblockA{\textit{Nanyang Technological University}\\
Singapore \\
hieuthi.luong@ntu.edu.sg}
\and
\IEEEauthorblockN{Haoyang Li\textnormal{*}}
\IEEEauthorblockA{\textit{Nanyang Technological University}\\
Singapore \\
li0078ng@e.ntu.edu.sg}
\and
\IEEEauthorblockN{Lin Zhang}
\IEEEauthorblockA{\textit{Brno University of Technology}\\
Czech Republic\\
zlin@ieee.org}
\linebreakand
\IEEEauthorblockN{Kong Aik Lee}
\IEEEauthorblockA{\textit{The Hong Kong Polytechnic
University}\\
Hong Kong SAR, China \\
kong-aik.lee@polyu.edu.hk}
\and
\IEEEauthorblockN{Eng Siong Chng}
\IEEEauthorblockA{
\textit{Nanyang Technological University}\\
Singapore \\
aseschng@ntu.edu.sg}
}

\maketitle

\begin{abstract}
Previous fake speech datasets were constructed from a defender's perspective to develop countermeasure (CM) systems without considering diverse motivations of attackers.
To better align with real-life scenarios, we created LlamaPartialSpoof, a 130-hour dataset that contains both fully and partially fake speech, using a large language model (LLM) and voice cloning technologies to evaluate the robustness of CMs.
By examining valuable information for both attackers and defenders, we identify several key vulnerabilities in current CM systems, which can be exploited to enhance attack success rates, including biases toward certain text-to-speech models or concatenation methods.
Our experimental results indicate that the current fake speech detection system struggle to generalize to unseen scenarios, achieving a best performance of 24.49\% equal error rate.

\end{abstract}

\begin{IEEEkeywords}
deepfake, dataset, fake speech detection, voice cloning, large language model.
\end{IEEEkeywords}

\section{Introduction}

The rapid growth in artificial intelligence brings unprecedented opportunities and risks \cite{masood2023deepfakes,akhtar2023multimodal} for society. 
The misuse of state-of-the-art (SOTA) voice cloning technology \cite{casanova2022yourtts,luong2020nautilus,du2024cosyvoice} to create and distribute disinformation is one of such risks.
ASVspoof\footnote{https://www.asvspoof.org/} is an initiative fosters the development of countermeasures (CMs) for this type of attacks \cite{Wu2015_asvspoof2015}. The latest edition, ASVspoof 5 \cite{Wang2024_ASVspoof5}, focused on large-scale crowdsourced speech and deepfakes.
Beside that, Many deepfake datasets \cite{frank2wavefake,reimao2019dataset,yaroshchuk2023open} were created using various text-to-speech (TTS) and voice conversion (VC) models for research purposes.
However, most were defender-centric created for the convenience of experimentation without much consideration on the attackers' perspective.
For example, some only contain fully synthetic utterances \cite{frank2wavefake,Wang2024_ASVspoof5} of arbitrary or original text \cite{muller2024mlaad,li2024cross} with limited amount of utterances \cite{yaroshchuk2023open} or attacking methods \cite{cai2023lav-df}.

From the attacker's point of view, it is desirable to replace just a small audio segment to change the utterance's meaning. The ``Partial Spoof'' (PS) scenario \cite{Zhang2021PartialSpoof}, which explored this type of attack, has attracted attentions recently.
The PartialSpoof \cite{Zhang2021PartialSpoof} database was created by randomly substituting real and fake audio segments. Its evaluation set included unknown spoofing methods, but it overlooked semantic consistency.
Meanwhile, the Half-truth dataset \cite{Yi2021halftruth} and Psynd database \cite{zhang2022psynd} addressed semantic consistency by replacing a single or several words. However, both of them only used a single TTS model.
In 2022, Cai \MakeLowercase{\textit{et al.}} released LAV-DF,
that used a rule-based system to replace a word with its antonym \cite{cai2023lav-df}, then in 2024, they published the AV-Deepfake1M dataset \cite{cai2023av}, which used ChatGPT\footnote{https://chat.openai.com} to alter sentences.
However, only two TTS models were used to create spoofed speech. 
In general, these datasets lack the quality and diversity to generalize to realistic scenarios.
Some had splicing artifacts that can be detected with spectral analysis methods \cite{Negroni2024_Analyzing}.

To create a fully and partially fake speech dataset based on the attackers' motivations, we present a generation pipeline that uses a large language model (LLM) \cite{dubey2024llama} and various TTS models. This pipeline can be used as a framework for deepfake creation and detection research.
We constructed the LlamaPartialSpoof dataset\footnote{https://github.com/hieuthi/LlamaPartialSpoof} using it and will release the dataset soon.
To ensure spoofed samples reflect the latest deepfake technologies, we selected several SOTA and widely accessible TTS systems. Lastly, we report our analyses on various aspects of creating and detecting partially fake speech, which is key for future development.

\begin{figure*}[t]

    \centering
    \fbox{\begin{minipage}{0.975\textwidth\fboxsep\fboxrule}
    \begin{subfigure}[b]{0.95\textwidth}
         \centering
         \includegraphics[width=\textwidth]{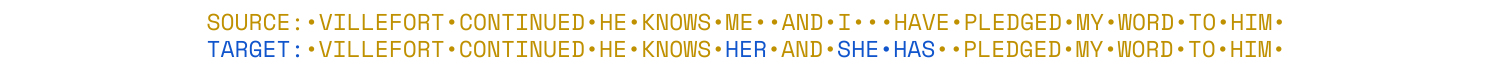} \\
         \vspace{-3mm}
         \caption{change one or two pronouns}
     \end{subfigure}
     
    \begin{subfigure}[b]{0.95\textwidth}
         \centering
         \includegraphics[width=\textwidth]{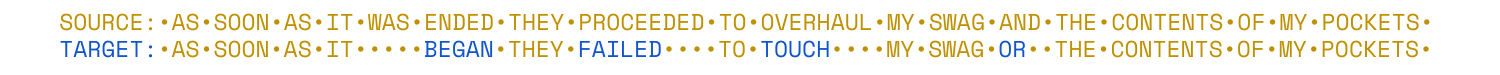} \\
         \vspace{-3mm}
         \caption{change the meaning to the opposite}
     \end{subfigure}
    \end{minipage}}
    \vspace{-1mm}
    \caption{Examples of sentences altered by the Llama 3 Instruct model.}
    \vspace{-5mm}
\end{figure*}

\section{Llama Partial Spoof Dataset}
\label{sec:llama-partial-spoof}

\subsection{Motivation}

To align with potential real-life threats, we created the LlamaPartialSpoof dataset, which includes both fully and partially fake speech utterances. Unlike the previous databases \cite{Yi2021halftruth,zhang2022partialspoof,zhang2022psynd}, ours was generated using a pipeline designed from attackers' perspective. It prioritizes content, intention, and quality of synthetic speech.
Since several studies have shown that training detection models on simulated data can improve performance\cite{zhang2022partialspoof,wang2024can,ge2024spoofing}, it is not necessary for the training data to be semantically correct.
Hence, we focus on providing a high-quality, realistic, and diverse evaluation set. The generation pipeline presented in this paper can be replicated to create training data under similar conditions.

\subsection{Generation Pipeline}

This section gives an
overview of our generation pipeline and the technical details of the LlamaPartialSpoof dataset.

\begin{table}[t]
    \caption{Number of fully fake and partially fake utterances generated from each TTS model. Word Error Rates (WERs) were calculated on 5,000 sentences.}
    \centering
    \scalebox{1.0}{
    \begin{tabular}{llrrrr}
        \hline \hline

         \multirow{2}{*}{ID} & \multirow{2}{*}{Model} & \multicolumn{2}{c}{\# of Utterances} & \multicolumn{2}{c}{WER (\%)} \\
         & & Full & Partial & Full & Partial\\ \hline
        TTS001 & LJ JETS & 5,577 & 5,387 & 2.28 & 3.59 \\
        TTS002 & YourTTS & 5,577 & 5,387 & 5.24 & 5.52 \\
        TTS003 & XTTS V2 & 5,577 & 5,384 & 2.29 & 3.42 \\
        TTS004 & GPT Sovits & 5,577 & 5,387 & 4.13 & 4.23 \\
        TTS005 & CosyVoice & 5,576 & 5,286 & 3.23 & 3.50 \\
        TTS006 & ElevenLab & 5,577 & 5,387 & \textbf{1.27} & \textbf{2.39} \\ \hline
    \end{tabular}}
    \label{tab:llamapartialspoof}
\vspace{-5mm}
\end{table}

\textbf{Data source:} We utilized the dev-clean subset of LibriTTS corpus \cite{zen19_interspeech}, comprising of 20 male and 20 female speakers, to construct our deep fake dataset.
Samples from both the dev-clean (5,736 utt.) and the test-clean (4,837 utt.) subsets were used as bona fide utterances.

\textbf{Transcript Manipulation:} Inspired by the AV-Deepfake1M dataset, we used the Llama-3-8B-Instruct model \cite{dubey2024llama} to automatically alter sentences. 
However, we directly asked the model to change a sentence instead of generating a series of replace, delete, and insert operations \cite{cai2023av} since that approach can limit its expressive capacity.
We prepared ten prompts, classified into two groups.
The first group changes parts of a sentence randomly: 1) one or two words; 2) several words; 3) one or two nouns; 4) one or two pronouns; 5) several nouns or pronouns.
The second alters a sentence slightly to change its meaning or conveyed emotion: 6) change the meaning to the opposite; 7) change the meaning to something else; 8) make it better; 9) make it simpler; 10) change the tone of the sentence.
Some requests failed due to the safety check of the Llama 3 model and we simply removed them.

\textbf{Fake Speech Generation:} Given the new fake sentences, we generated fully synthetic utterances using either commercial or open source TTS systems. Five out of six systems were capable of cloning voices and the last one was a single speaker system with the voice of a female speaker. Section \ref{subsec:tts-models} describes each model in more details.

\textbf{Partial Spoof Creation:}
Using the original bona fide and fully fake utterances, we extracted word alignment information with the Montreal Forced Aligner \cite{mcauliffe17_interspeech} and leveraged it to create partially spoofed utterances by combining them with minimal editing.
To concatenate real and fake segments, we can cut and paste their waveform, using overlap-add \cite{zhang2022partialspoof}, or crossfading method.
From an audio engineer's perspective, crossfading is more versatile as it has more parameters for customization.
Specifically, we randomly assigned the length of the overlap between 30 and 80 ms then used one of these five fading functions: linear, quarter or half of sine wave, logarithmic, and inverted parabola.
Loudness was normalize to increase the compatibility.

\textbf{Post-processing:} Both the bona fide and the fake utterances were downsampled to 16 kHz and assigned a random peak audio level from -0.01 dBFS to -10 dBFS.

\begin{figure}[t]
         \centering

    \begin{subfigure}[b]{0.48\columnwidth}
         \includegraphics[width=\textwidth]{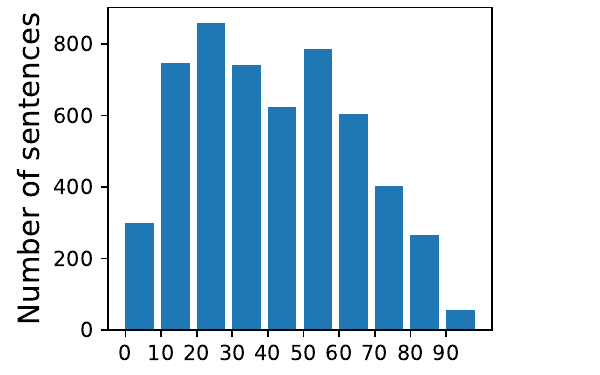}
         \vspace{-6mm}
         \caption{percentage of fake words}
     \end{subfigure}%
    \begin{subfigure}[b]{0.48\columnwidth}
         \centering
         \includegraphics[width=\textwidth]{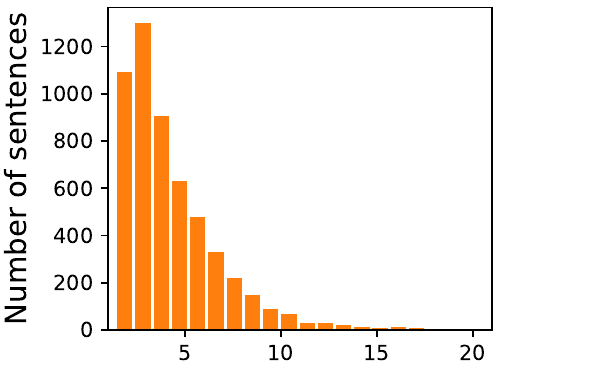}
         \vspace{-6mm}
         \caption{number of fake segments}
    \end{subfigure}
    \vspace{-2mm}
    \caption{Histograms illustrate number of fake sentences that have a given percentage of fake words or fake segments.}
    \label{fig:dataset_analysis}
    \vspace{-5mm}
\end{figure}

\begin{table*}[t!]
    \caption{Datasets used for training and evaluating.}
    \centering
    \scalebox{1.0}{
    \begin{tabular}{lllllrrrr}
        \hline \hline
         \multirow{2}{*}{Ref.}& \multirow{2}{*}{Dataset} & \multirow{2}{*}{Datasource} & Language / & Concatenating & \multirow{2}{*}{\#Attacks} & \multicolumn{3}{c}{Number of Utterances} \\
         &   &  & Sematically & method & & Bona fide & Full Fake & Partial Fake \\ \hline
        \multirow{3}{*}{\cite{Zhang2021PartialSpoof}} & PartialSpoof-train & \multirow{3}{*}{ASVspoof 2019 LA \cite{wang2020asvspoof}} & \multirow{3}{*}{EN/No} & \multirow{3}{*}{\tabincell{l}{cross-correlation \\ + overlap-add}} & 6 & 2,580 & 0 & 22,800 \\
        & PartialSpoof-dev & & & & 6 & 2,548 & 0 & 22,296 \\
        & PartialSpoof-eval & & &&  13& 7,355 & 0 & 63,882 \\ \hline

        \multirow{2}{*}{\cite{Yi2021halftruth}} & Half-truth-train & \multirow{2}{*}{AISHELL-3 \cite{shi21c_interspeech}} & \multirow{2}{*}{CN/Yes} & \multirow{2}{*}{unspecified} & \multirow{2}{*}{1}  & 26,554 & 1,185 & 25,354 \\
        & Half-truth-dev & & & & & 8,914 & 430 & 8,480 \\\hline

        \multirow{2}{*}{\cite{cai2023av}} & Deepfake1M-train & \multirow{2}{*}{Voxceleb2 \cite{chung18b_interspeech}} & \multirow{2}{*}{EN/Yes} & \multirow{2}{*}{unspecified} & \multirow{2}{*}{2}  & 186,666 & 0 & 186,344 \\
        &Deepfake1M-val & & & & & 14,235 & 0 & 14,515 \\\hline
        
        \multirow{2}{*}{\cite{Wang2024_ASVspoof5}} & ASVSpoof5-train & \multirow{2}{*}{Multilingual Librispeech \cite{pratap20_interspeech}} &  \multirow{2}{*}{EN/Yes} & \multirow{2}{*}{N/A} & 8 & 18,797 & 163,560 & 0 \\ 
        &ASVSpoof5-dev & & & & 8 & 31,334 & 109,616 & 0 \\ \hline

        Ours & LlamaPartialSpoof & LibriTTS \cite{zen19_interspeech} & EN/Yes & Crossfade & 6 & 10,573 & 33,461 & 32,194 \\ \hline
    \end{tabular}}
    \label{tab:datasets}
\vspace{-3mm}
\end{table*}

\subsection{Text-to-speech Models}
\label{subsec:tts-models}
We chose five open-source models and one commercial service based on their quality, popularity, and their differences from each other to generate fake utterances.

\textbf{LJ JETS:} This single female speaker TTS model \cite{lim22_interspeech} was trained on the 24-hour LJ Speech dataset\footnote{https://keithito.com/LJ-Speech-Dataset/}. It\footnote{https://github.com/espnet/espnet/tree/master/egs2/ljspeech/tts1} is an end-to-end model that jointly training the FastSpeech 2 feature generator \cite{ren2021fastspeech} and the HiFi-GAN neural vocoder \cite{kong2020hifi}. We used it as a non-voice-cloning reference.

\textbf{YourTTS:} We used the multilingual YourTTS model\footnote{https://github.com/coqui-ai/TTS/} and 3--7 seconds of speech to clone voices.
It is a modified of VITS \cite{kim2021conditional}, which utilizes a transformer-based encoder for text encoding and a speaker encoder for speaker embedding \cite{casanova2022yourtts}.

\textbf{XTTS V2:}
We used the XTTS V2 system\footnote{https://huggingface.co/coqui/XTTS-v2} and 3--7 seconds of speech to clone voices. The model was trained to predict VQ-VAE audio codes of target speech using a decoder-only transformer. Speech was created using a HiFi-GAN vocoder and the latent space of the transformer \cite{casanova2024xtts}. 

\textbf{GPT-SoVITS:}
We finetuned the GPT-SoVITS model\footnote{https://huggingface.co/lj1995/GPT-SoVITS/tree/main} using the dev-clean set to enhance speaker similarity, then clone voices using a single utterance of the speaker.
The system used discrete semantic tokens \cite{hsu2021hubert} to train a text-to-semantic model and a VITS decoder to synthesize speech.

\textbf{CosyVoice:}
The pretrained CosyVoice-300M model\footnote{https://github.com/FunAudioLLM/CosyVoice} uses text and acoustic prompts to predict semantic tokens of the target speech, which was used to generate mel-spectrogram \cite{du2024cosyvoice}.
During inference, we used the fake sentence and the real utterance to generate the fake utterance.

\textbf{ElevenLab:} We used ElevenLab's voice cloning service\footnote{https://elevenlabs.io} and 5 utterances per speaker to clone their voices then generating fully fake utterances. Since ElevenLab is a commercial service, it is a black-box system.

\subsection{Dataset Analysis}

We created 5,577 spoofed sentences out of 5,736 originals after checking outputs of the Llama model.
Among those, 5,387 sentences were used to create partially fake utterances, excluding 190 sentences only had delete operations.
Table \ref{tab:llamapartialspoof} shows the number of fully and partially fake utterances created with each TTS model\footnote{TTS005 has fewer samples due to failed alignment in some cases}.
We calculated word error rate (WER) on 5,000 sentences using Whisper large model\footnote{https://huggingface.co/openai/whisper-large}. 
Overall, partial subsets had higher WER than their full counterparts with ElevenLab got the best results.
Figure \ref{fig:dataset_analysis} shows the distributions of fake sentences across different percentage of fake words or different number of fake segments. Comparing with the AV-Deepfake1M dataset, our partial subset has more utterances with more than two fake segments. The shortest fake segment is 0.03 s, while the longest is 24.72 s, with a mean of 0.95 s and a standard deviation of 0.85 s.

\begin{table}[t]
    \captionsetup[subtable]{position = top}
    \caption{Cross-dataset EERs. Gray cells indicate the train set includes data from the same corpus as the evaluation. Bold values indicate the lowest EER (in each column) of the particular evaluation set excluding all the gray cells.}
    \centering
    \begin{subtable}{1.0\columnwidth}
        \centering
        \caption{Utterance-based EERs (\%)}
        \scalebox{1.0}{
        \begin{tabular}{l|rrrrr}
            \hline \hline
             Train Set & \multicolumn{5}{c}{Evaluation Set} \\ \cline{2-6}
             & ps-eval & had-dev & 1m-val & asv5-dev & ours \\ \hline
            \texttt{A}. ps-train & \cellcolor{gray!25} 1.48 & 40.40 & \textbf{39.37} & \textbf{5.14} & \textbf{24.49} \\
            \texttt{B}. had-train & 49.51 & \cellcolor{gray!25} 0.02 & 49.81 & 42.24 & 57.18 \\
            \texttt{C}. 1m-train & 51.08 & 56.29 & \cellcolor{gray!25} 0.02 & 51.41 & 51.63 \\\hline
            \texttt{D}. asv5-train & \textbf{30.73} & \textbf{33.22} & 43.99 & \cellcolor{gray!25} 4.50 & 34.21 \\ \hline
            \texttt{A+B+C} & \cellcolor{gray!25} 1.19 & \cellcolor{gray!25} 0.06 & \cellcolor{gray!25} 0.04 & 14.07 & 31.87 \\\hline
        \end{tabular}}
    \end{subtable}
    \vspace{-1mm}
    
    \begin{subtable}{1.0\columnwidth}
    \centering
        \caption{20-ms Segment-based EERs (\%)}
        \vspace{-1mm}
        \begin{tabular}{l|rrrrr}
            \hline \hline
             Train Set & \multicolumn{5}{c}{Evaluation Set} \\ \cline{2-6}
             & ps-eval & had-dev & 1m-val & asv5-dev & ours \\ \hline
            \texttt{A}. ps-train & \cellcolor{gray!25} 13.70 & 47.41 & \textbf{20.59} & \textbf{5.81} & 47.49 \\
            \texttt{B}. had-train & 43.97 & \cellcolor{gray!25} 0.12 & 28.61 & 35.56 & 59.36 \\
            \texttt{C}. 1m-train & 49.84 & 40.76 & \cellcolor{gray!25} 0.18 & 60.60 & 51.16 \\\hline
            \texttt{D}. asv5-train & \textbf{34.20} & \textbf{37.74} & 34.99 & \cellcolor{gray!25} 2.99 & \textbf{38.50} \\ \hline
            \texttt{A+B+C} & \cellcolor{gray!25} 13.59 & \cellcolor{gray!25} 0.08 & \cellcolor{gray!25} 0.19 & 26.46 & 50.16 \\\hline
        \end{tabular}
    \end{subtable}

    \label{tab:cross-dataset}
\vspace{-3mm}
\end{table}

\section{Experiments}
\label{sec:experiments}

In this research, we did not propose a novel methodology for fake speech detection, instead focusing on evaluating the performance of existing systems on our comprehensive dataset, which yielded valuable insights for advancements of robust detection systems. We leveraged a diverse range of publicly available deep fake datasets to train our model.

\begin{table*}[t!]
    \caption{Utterance-based EER (\%) calculated on each TTS model subsets of LlamaPartialSpoof. The results were calculated on 5,000 sentences. Bold values indicate the lowest EER (in each column) of the particular evaluation subset.}
    \centering
    \scalebox{1.0}{

    \begin{tabular}{l|rr|rr|rr|rr|rr|rr|rr}
        \hline \hline
         Train Set & \multicolumn{2}{c|}{LJ JETS} & \multicolumn{2}{c|}{YourTTS} & \multicolumn{2}{c|}{XTTS V2} & \multicolumn{2}{c|}{GPT-SoVITS} & \multicolumn{2}{c|}{CosyVoice} & \multicolumn{2}{c|}{ElevenLab} & \multicolumn{2}{c}{All} \\ \cline{2-15}
          & Full & Partial & Full & Partial & Full & Partial & Full & Partial & Full & Partial & Full & Partial & Full & Partial \\ \hline 
        \texttt{A}. ps-train & 47.31 & \textbf{17.14} & 20.25 & 10.72 & 27.11 & \textbf{14.90} & 18.47 & \textbf{13.50} & \textbf{30.62} & \textbf{16.92} & 46.13 & \textbf{18.37} & \textbf{32.35} & \textbf{15.33} \\
        \texttt{B}. had-train & \textbf{44.93} & 36.02 & 66.55 & 49.65 & 70.58 & 52.41 & 68.21 & 53.56 & 66.37 & 47.50 & 69.35 & 47.87 & 65.21 & 47.54 \\
        \texttt{C}. 1m-train & 55.23 & 48.99 & 54.12 & 48.26 & 55.68 & 49.58 & 54.79 & 50.09 & 54.05 & 48.22 & 51.21 & 47.36 & 54.17 & 48.90 \\\hline
        \texttt{D}. asv5-train & 44.98 & 44.10 & \textbf{0.63} & \textbf{9.03} & \textbf{24.25} & 26.33 & \textbf{6.60} & 22.63 & 40.17 & 37.64 & 48.41 & 45.29 & 34.14 & 34.52 \\ \hline
        \texttt{A+B+C} & 50.11 & 27.84 & 18.39 & 17.24 & 32.61 & 20.83 & 40.07 & 24.80 & 47.88 & 27.35 & \textbf{38.75} & 26.57 & 39.00 & 24.55 \\\hline          

    \end{tabular}}
    \label{tab:tts_eer}
\vspace{-5mm}
\end{table*}

\subsection{Fake Speech Detection Model}

The multi-resolution model \cite{zhang2022partialspoof} was used for all of our experiments. It was trained to detect fake segments at 20, 40, 80, 160, 320, and 640 ms resolutions and at utterance-level. Only the results at the utterance level and the 20-ms resolution was presented to keep it simple. The configuration was the same as in the original paper but the length of training sequences.
We used 9.6-second random segments from training samples for training, to facilitate experimentation across several datasets.
In evaluation, we segmented long utterances into multiple 9.6-second parts then combined their results. For the utterance-level score, the maximum value of the spoof class among the parts was used.

\subsection{Datasets}

Several datasets created by different research groups were used in the experiments (Table \ref{tab:datasets}).
The train (ps-train) and evaluation (ps-eval) set of PartialSpoof were used for training and evaluating. Similarly, we used the Half-truth train (had-train) and development (had-dev) set since its test set does not contain bona fide samples.
About the AV-Deepfake1M corpus, we extracted audios from real-video-real-audio and fake-video-fake-audio samples of the train (1m-train) and validation (1m-val) subset for the experiments since labels of the test set is not available.
The ASVspoof5 train (asv5-train) and development (asv5-dev) set were included to measure the performance of CMs against a fully spoofed speech.
The LlamaPartialSpoof dataset (ours) was used for evaluation.

\subsection{Metrics}

For evaluations, we used the utterance-based Equal Error Rate (EER) and a 20-ms segment-based EER as metrics. 
The utterance-based metric is suitable for assessing the performance of the fake speech detection task, whereas segment-based ones are better for the fake segment localization task.

\subsection{Results and Analysis}

\begin{table}[t]
    \caption{Utterance-based EER (\%) calculated on subsets of LlamaPartialSpoof. Gray cells indicate the lowest EER (in each row) produced by a particular model.}
    \centering
    \begin{subtable}{1.0\columnwidth}
        \centering
        \caption{partially fake subsets by the percentage of fake words}
        \label{tab:fakeword_eer}

        \scalebox{1.0}{
        \begin{tabular}{l|rrrrr}
            \hline \hline
             Train Set & \multicolumn{5}{c}{subsets by percentage of fake words} \\ \cline{2-6}
             & 0--20\% & 20--40\% & 40--60\% & 60--80\% & 80--100\% \\ \hline
            \texttt{A}. ps-train & 22.80 & 13.51 & \cellcolor{gray!25}11.12 & 12.76 & 19.42 \\
            \texttt{B}. had-train & \cellcolor{gray!25}44.37 & 47.59 & 47.56 & 47.84 & 52.16 \\
            \texttt{C}. 1m-train & 50.43 & \cellcolor{gray!25}47.41 & 48.07 & 50.10 & 51.33 \\\hline
            \texttt{D}. asv5-train & 41.03 & 36.05 & \cellcolor{gray!25}27.14 & 27.15 & 31.66 \\ \hline
            \texttt{A+B+C} & 28.78 & 24.21 & \cellcolor{gray!25}21.54 & 22.68 & 26.68 \\\hline
        \end{tabular}}
    \end{subtable}

    \vspace{2mm}
    \begin{subtable}{1.0\columnwidth}
    \centering
    \caption{fully fake or partially fake subsets}
    \label{tab:concatenation_eer}

    \scalebox{1.0}{
    \begin{tabular}{l|r|rrrr}
        \hline \hline
         Train Set & Fully fake & \multicolumn{3}{c}{Partially fake} \\ \cline{3-5}
         & & Crossfade & Cut/Paste & Overlap-Add  \\ \hline
        \texttt{A}. ps-train & 32.07 & 15.21 & 15.37 & \cellcolor{gray!25}13.89 \\
        \texttt{B}. had-train & 64.78 & 47.24 & \cellcolor{gray!25}32.72 & 37.89 \\
        \texttt{C}. 1m-train & 53.66 & \cellcolor{gray!25}49.04 & 54.90 & 54.06 \\\hline
        \texttt{D}. asv5-train & 34.17 & 34.26 & 35.11 & \cellcolor{gray!25}28.56 \\ \hline
        \texttt{A+B+C} & 38.70 & 24.54 & \cellcolor{gray!25}21.83 & 24.21 \\\hline
    \end{tabular}}
    \end{subtable}
\vspace{-5mm}
\end{table}

\subsubsection{Cross-dataset fake speech detection}
Table \ref{tab:cross-dataset} presents the utterance-based and 20-ms segment-based EER when training and evaluating across datasets. The model A+B+C was trained using the combination of three partially fake training sets.
As expected, both types of EERs are every low when the train and evaluation set come from the same source, but they are very high otherwise.
Model A trained with ps-train had the best utterance-based score while model D trained with asv5-train had the best segment-based when tested with the LlamaPartialSpoof dataset.
We believe The diverse range of attacks in the training sets helped these models generalize better to unseen ones.
The degradation performances when training and testing across datasets highlight the need to develop better strategies for domain adaption and generalization.

\subsubsection{EERs of fully and partially fake subsets}
Table \ref{tab:tts_eer} shows the EER evaluating on the partial and full subsets generated by different TTS models. Generally speaking, model A is better at detecting partially fake than fully fake while it is the opposite for model D due to the nature of their training data.
LJ JETS had the highest lowest-EER for the fully fake scenario while it is ElevenLab for the partially fake.
Interestingly, partially fake samples of LJ JETS got the 2nd highest EER which suggests the CMs did not take speaker consistency into account.
Model D, trained with ASVspoof5 data, can detect YourTTS fully fakes samples quite well as its data also includes samples created with YourTTS \cite{Wang2024_ASVspoof5}.
Since model A, trained on the PartialSpoof dataset, had the best performance on detecting partial subsets, it reiterates that semantically ignorant data can be used to train partially fake speech detection model.

\subsubsection{EERs by the percentage of fake words}
Table \ref{tab:fakeword_eer} shows EERs on five partially fake subsets, each contains sentences with a particular percentage of fake words. We noticed most models had higher EERs on 0--20\% and 80--100\% subsets, and low EERs on the rest. Intuitively, this suggests CM systems are more susceptible to a real (synthetic) utterance that contains an arbitrarily short synthetic (real) segment.

\subsubsection{Effects of concatenating methods}
Table \ref{tab:concatenation_eer} shows the EER based on utterance calculated with fully or partially fake samples created with different methods.
Model A trained with PartialSpoof, which used cross-correlation and overlap-add, is better at detecting samples created with the same method.
Interestingly, model B and C are better at detecting samples created with cut/paste and crossfade respectively, but their EERs are still very high.
Model D favors samples created with overlap-add even though its training data does not contain partially fake samples.
This highlights concatenation methods as an aspect of interest for both attackers and defenders.

\section{Conclusion}
In this paper, we show how attackers can utilize LLMs and voice cloning technologies to create mass disinformation media.
Researchers can also use this framework to conduct experiments on deepfake creation and detection across transcripts and audio \cite{akhtar2023multimodal}.
Our 130-hour LlamaPartialSpoof dataset is an initiative to create a large-scale out-of-domain evaluation set for CM research and development.
Several aspects of partially fake speech creation were investigated, but the overall methodology was still very simple.
For the future work, researchers can focus on improving performances of CM systems on these scenarios or more advanced ones \cite{cuccovillo2022open} such as in noisy \cite{li2024cross} or reverberant environments \cite{luong2024room}.


\bibliographystyle{IEEEbib}

\bibliography{refs}

\end{document}